# A Survey on Technological Trends to Enhance Spectrum Efficiency in 6G Communications

[a]Sridhar Iyer, [b]Anita Patil, [c]Shilpa Bhairanatti, [d]Soumya Halagatti, and [e]Rahul Jashvantbhai Pandya

[a,*]Dept. of ECE, KLE Dr. M.S. Sheshgiri College of Engineering and Technology, Udyambag, Belagavi, Karnataka, India-590008 (E-mail: sridhariyer1983@klescet.ac.in)

[b,c,d]Dept. of ECE, S.G. Balekundri Institute of Technology, ShivabasavaNagar, Belagavi, Karnataka, India-590010

[e]Dept. of EE, Indian Institute of Technology-Dharwad, WALMI Campus, PB Road, near High Court, Karnataka, India-580011

**Abstract:**

The research community has already identified that, by 2030, 5G networks will reach the capacity limits, and hence, will be inadequate to support next generation bandwidth-hungry, ubiquitous, intelligent services, and applications. Therefore, in view of sustaining the competitive edge of wireless technology and stratifying the next decade's communication requirements both, industry and research community have already begun conceptualizing the 6G technology.

This article presents a detailed survey on the recent technological trends which address the capacity issues and enhance the spectrum-efficiency in 6G Communications. We present these trends in detail and then identify the challenges that need solutions before the practical deployment to realize 6G communications. Our survey article attempts to significantly contribute to initiating future research directions in the area of spectrum-efficiency in 6G communications.

## I. Introduction

The exponential increase in the traffic volume can be associated with the fast-paced development of next-generation technologies catering to the needs of emerging applications, such as augmented reality (AR) and virtual reality (VR), Internet of Everything (IoE), artificial intelligence (AI), robotics and automation, etc. According to the report presented in [1], the overall global traffic, which was estimated to be approximately 7.4 EB/month in the year 2010, is predicted to increase to approximately 5000 EB/month in the year 2030. The aforementioned statistic only suggests that, even though there has been a tremendous evolution in the information and communications technology, there still exists a desperate requirement to improve the communication systems and the devices. For the past few decades, the wireless communication technology community has showcased the ability to keep pace with the growing demands, simultaneously ensuring the inclusion of the latest trends in the standards, and ensuring multiple breakthroughs in the mobile networks. With the human society advancing towards complete automation and remote management, it is only a matter of time before very high capacity systems ensuring reliability and cost/power-efficiency become the need of the hour. In this regard, the 5G technology, which is expected to be completely deployed and operational worldwide by the year 2020, has been envisioned to provide the necessary solutions [2-6]. In fact, according to the survey in [7], for the 5G mobile communication system, the standard for the non-standalone (NSA) version was completed in the year 2017, and a standalone (SA) version was then finalized towards the end of the year 2018 by the 3rd Generation Partnership Project (3GPP).

The 5G technology has been considered as the catalyst to the start of a complete digitally equipped human society simultaneously attaining substantial breakthroughs concerning latency, capacity, mobility, and the number of connected devices [7]. The 5G system has been envisioned to provide enhanced service quality in comparison to the previous wireless generations, such as 4G [2, 8, 9]. To provision high capacity demanding applications, 5G standards include the implementation of (i) multiple high-frequency spectra, such as millimeter–waves (mm) and optical frequency, (ii) advanced frequency sharing, utilizing and managing methods, (iii) a mixed frequency spectrum band comprising of both, the licensed and the unlicensed bands, (iv) increased granularity in the distribution of network functions, (v) dynamic configuration of the network infrastructure through network function virtualization (NFV), (vi) virtual network functions (VNFs) over a "white box" hardware resulting in cost savings, (viii) separate quality of service levels for specific applications over a single physical network (i.e., network slicing) [10, 11]. Besides, the 5G systems will also be able to support multiple services, such as emergency location, enhanced mobile broadband (eMBB), ultra-reliable and low-latency communications (uRLLC), and massive machine-type communications (mMTC), etc. [11, 12]. It has also been reported that the 5G technology in the (i) initial stage will be in a position to enable high capacity and low-latency driven applications, such as real-time closed-loop robotic control, video-driven machine-human interaction, and AR/VR, and (ii) evolving stage will support even larger capacity and lower latency (i.e., ultra-high-speed low-latency) seeking applications, such as sharing and updating high-resolution maps in real-time for control of autonomous vehicles [13].

The 5G standardization is already complete, and currently, the system is being deployed worldwide. The 5G technology promises higher reliability, improved power-efficiency, and easier reconfiguration of the network for deploying additional services [2]. Hence, in comparison to the existing technology, the 5G technology will provide significant improvements [14]; however, a major drawback of the 5G technology is that it does not standardize the convergence of various functions such as communication, sensing, security, computing, etc. [15]. With automated and intelligent service requirements, the various 5G functionalities will need to converge for provisioning the IoE scenario, and with a fast-paced move of the society towards a fully automated and intelligent network scenario, the 5G technology will also need to be upgraded in view of provisioning enhanced services with fully immersive experiences [16]. The aforementioned points towards the requirement of high bandwidth and intelligence within the 6G technology which are detailed next.

Firstly, with the increased traffic volume and the number of connected devices, the bandwidth requirement will also rapidly grow, which will pose immense challenges to accomplish enhanced spatial- and spectral-efficiency. Next-generation applications, such as holographic videos, VR, and ubiquitous connectivity, will require very high bandwidth, which cannot be supported by the mm-wave spectrum (30–100 GHz) used in the 5G technology. Therefore, to obtain a broader radio spectrum bandwidth and extremely wideband channels with tens of GHz-wide bandwidth, TeraHertz (THz) communication (i.e., 0.1-10 THz) looks a promising option [13, 16]. In fact, in the year 2019, the Federal Communications Commission (FCC) designated the range of 95 GHz to 3 THz spectra for practical use and unlicensed applications in view of encouraging the development of 6G communications targeting the Tbps data rate [17]. The report in [16] also mentions the use of up to 3 THz by the 6G technology-based advanced applications. Apart from the bandwidth enhancement, THz communication is also expected to realize other major applications such as provisioning high-precision positioning capability owing to the (i) accurate ranging between the transmitter and the receiver by the wideband THz communication, (ii) precise line of sight (LoS) between the transmitter and the receiver, and (iii) pin-point sharp beams which are steeped in both, azimuth and elevation [18, 19]. However, as reported in [20], owing to the propagation losses, THz communication will be limited to high bit-rate short-range communications. Lastly, as detailed in [21], for 6G, to provide the increased capacity required by the advanced bandwidth-hungry applications, there will be a transition from the radio frequency to sub-THz, and visible light communication (VLC) (i.e., 400 THz-800 THz). Overall, it can be concluded that there are many apparent benefits to implement the THz communication for 6G; however, before the practical deployments, it is mandated to provide solutions for the many existing technical challenges, which are detailed in the later sections of this article.

Secondly, in view of ensuring intelligence within 6G communications, AI technology will be mandated, which will reside in the new local cloud environments and will help to create a plethora of new applications via the use of sensors that have already been embedded within every aspect of the human life. The research community has already foreseen the multiple techniques and applications which will leverage the AI technologies in wireless networks [22, 23]. Mobile edge computing has been identified as the technology which will transform the 5G communication networks into distributed cloud computing platforms [13]. However, the 5G networks will not realize the unlimited potential of the AI methods since these were not accounted for in the 5G architecture's starting design; instead, only became a part of the design with time [24, 25]. Therefore, at the maximum, the 5G systems can use the AI technology to optimize the network's legacy architecture [26]. However, for 6G communications, AI technologies hold the massive potential to improve network performance and minimize the overall network cost (both CapEx and OpEx) [27]. Further, Machine Learning (ML) and Neural Networks (NNs) provide the opportunities to solve the link and the system-level problems in 6G communications simultaneously including features such as, self-configuration, aggregation, opportunistic set-up, and context awareness [19, 20]. As an example, it has been reported in [24] that an AI-enabled 6G network utilizing the ML algorithms can realize radio signalling and maximum cognition to the intelligent radio transmission. It must also be pointed out that ML has already attracted much attention from both, the industry and the academia for the implementation of various domains of the 5G technology, including applications at the (i) physical layer such as, coding and estimation of the channel, (ii) MAC layer such as, multiple access, (iii) network layer such as, resource allocation and error correction, etc. [28-30]. In conjunction with edge computing, AI technologies have paved the way to new-age applications, such as healthcare, simultaneously ensuring the minimization of the costs and improvement of the Quality of Experience [13, 26]. Also, concerning the spectrum resource management, most of the encountered algorithms are combinatorial, and hence, the existing optimal exhaustive-search based algorithms cannot be implemented in practice. Therefore, the ML-enabled algorithms will be mandated in view of outperforming the current sub-optimal approaches. As an example, the reinforcement learning algorithms and the transfer learning methods can be used which will learn, adapt, and optimize in response to the varying conditions over time and will be expected to be sufficient for performing the resource management tasks with minimal supervision. This will, in turn, include the investigation of power control, beamforming in the massive Multiple Input Multiple Output (MIMO) and cell-free environments, predictive scheduling and resource allocation, etc.

Further, as stated earlier, the exponential growth in the demand for high data rates and the densification in the mobile devices motivates researchers to explore the THz band supporting the aggregated data rate of Tbps and user-centric data rate of Gbps. Therefore, it is crucial to enhance the spectrum-efficiency in the THz communications. The recent advances have shown a significant improvement in the spectrum-efficiency when ML techniques have been applied in resource allocation and medium access control [31, 32]. The authors in [33] have put forth the necessity of ML techniques in dynamic spectrum management of 6G communications to enhance spectrum-efficiency and have also summarised several ML techniques, which can be applied in dynamic spectrum sensing such as, supervised ML, unsupervised ML, K-nearest neighbor, support vector machine (SVM), K-means, Gaussian Mixture Model, etc. Moreover, in deep learning, the convolutional neural networks (CNN), recurrent neural network (RNN), and deep reinforcement learning (DRL) have been identified as the candidates for spectrum sensing and dynamic spectrum management to improve the spectrum-efficiency in 6G communications [33]. The authors in [34] have presented the efficient spectrum usage techniques for THz communication; the new spectrum can carry very high data rates; however, it is challenging to sustain the increased path losses and the impact of other impairments during the transmission. Therefore, it is concluded that efficient spectrum management techniques, new radio access techniques, and intelligent automation using ML will enhance the spectrum-efficiency in THz communication. Moreover, reconfigurable front ends with dynamic spectrum access smart sensing will significantly improve the spectrum-efficiency. In [35], the authors have presented an extensive study of the deep learning methods applied for spectrum monitoring, modulation recognition, and spectrum assignment in cognitive radio to improve the spectrum-efficiency. The authors in [36] have summarized various ML oriented techniques for opportunistic spectrum access and spectrum sharing in view of enhancing the spectrum-efficiency for 5G and beyond communications. In [37], the authors have presented the multi-agents-based deep learning techniques in resource allocation for D2D communication for improving the spectrum-efficiency. The authors in [38] have demonstrated the K-nearest neighbour learning based spectrum sensing for the cognitive radio network for improving the spectrum-efficiency.

Hence, it is clear that over the next decade, (i) in view of fulfilling the vision of a ubiquitous intelligent network, the existing 5G technology must be upgraded such that it includes AI within the network architecture comprehensively, thereby following an AI-driven approach where, intelligence will be an endogenous characteristic of the architecture, and (ii) the 5G technology will also reach the capacity limits which will require larger bandwidth, wider spectrum, and higher spectrum-efficiency. The two aforementioned points motivate the 5G successor technology drive, namely, the 6G technology [13, 16, 19-22, 27, 28].

Several studies have surveyed the 6G evolution and application in detail [7, 39, 40, 41]. A detail of the existing surveys and tutorials, including their main focus, is shown in Table 1. In comparison to these existing surveys, our survey discusses the state-of-the-art advances with a focus on the communication and spectrum issues rather than many other technologies and applications of the 6G technology. Further, towards the end of the survey, we also present numerous novel open research challenges and future research directions. In the opinion of the authors, the main contributions of this survey are as follows:

- We explore and discuss state-of-the-art advances made towards enabling the 6G systems.
- We devise a taxonomy of the 6G wireless systems based on the crucial enablers, use cases, emerging AI/ML schemes, and communication, networking and computing technologies.
- We discuss several open research challenges and their possible solutions with an outlook towards the future research on 6G communications.

The rest of the survey is organized as follows: Section II presents the various applications of the 6G technology. In Section III, we present the various technological trends which this survey article focuses on. Section IV presents the various global research activities which have been conducted concerning the 6G technology. The open research challenges with potential directions are presented in Section V. Finally, the conclusion is presented in Section VI.

Table 1. Summary of the existing surveys and tutorials.

| Reference | Recent Advances | Key Enabling Techniques | Use Cases | Remarks |
|---|---|---|---|---|
| [13] | Collection and Distribution on AI based IoT Communication | Block chain,SDN/NFB software control | 1.AI Based Edge networking for real time operation.<br>2.Optimization and Automation in 6G networking. | Dynamic networking is proposed based on artificial intelligence. |
| [18] | THz communication | Machine learning | 1.Satellite Integrated network<br>2.Connected Intelligence<br>3.Seemless Integration of wireless information and Energy Transfer<br>4.Ubiquitous super 3D connectivity. | The parameters like Free apace optical network,3 dimensional networking, quantum communication, cell free communication integration of wireless information and energy transfer, integrated sensing and communication, integrated access-backhaul networks, dynamic network slicing, holographic beamforming are analysed. |
| [20] | Sub-THz Communication | AI/ML, Holography | 1.Cluster size ,overlap between clusters, number of collaborating Base stations, trade-off between performance and complexity. | Impact of 6G on transportation industry. |
| [21] | Physical layer modelling, LIDAR communication. | Single carrier modulation, constant and near constant envelope modulation. | Efficient Spectrum utilization (90-200ghz) | Increase in the wireless network capacity is observed. |
| [22] | KNN | ML based Data Regression | 1.Deployment of models for processing and training high velocity data. | Supervised and unsupervised ML techniques are used to analyse IOT smart data. |
| [23] | Linear detection-QuaDRiGa | M-MIMO | 1.Rubustness of conjugate gradient. | Suitability for small set of multipliers is proved. |
| [24] | | Candidate technology | Relationship of candidate technology with AI | Effectiveness of AI to manage cellular network resources with good performance is studied. |
| [25] | RMCS (Robust Mobile crowd Sensing) | Deep learning<br>Edge computing | Implementation of robust mobile crowd sensing | Use of edge computing to solve issues related to latency, traffic load |
| [26] | KCCH | Edge learning | Knowledge centric connected healthcare. | Local processing of health supervision data by KCCH.<br>Edge learning helped the patient to choose good guardian, simulation results confirmed the satisfactory performance in data accuracy. |
| [29] | Intelligent adaptive learning | ML | Development of smart radio terminals | Inference control and power adjustment protocols are proposed to achieve efficient learning. |
| [30] | ML | ML | Access of large data from network and subscribers: Cost effective operational designed are proposed. | Discussion on proactive, self-aware, self-adaptive, predictive network has been done. |
| [40] | 1.Transformation of cognitive radio to Intelligent radio.<br>2.Reconfigurable Intelligent Surfaces | - | 1.High quality communication<br>2.World-wide connectivity | Features like Self aggregation, self-configuration, context awareness, opportunistic set up are analysed |
| [46] | Quantum cascade lasers. | - | Wide Frequency tuneable, compact, bright source of THz radiation:<br>Nitrous oxide is used to illustrate wide tunability. | Gas phase molecular laser based rotational population inversions optically pumped by quantum cascade lasers are suggested. |
| [49] | Quantum computing | Novel QC assisted and QML | Framework design to articulate 6G network infrastructure | Intelligent proactive catching ,MEC ,multi objective routing optimization research allocation, big data analytics, Interofidelity harmonization, secure link assurance are thoroughly discussed and recommended |
| [50] | Green 6G Network | AI, THz communication | - | Green 6G network which delivers high quality of Service and efficiency by offering fast and integrated service to ubiquitous users with on demand personality to anywhere any time |
| [52] | hybrid –electronic and photonic communication | THz integrated communication | Creation of High power levels using III-V based semiconductors high electron mobility transistors, GaAs based shottkey diodes. | Silicon based integrated technology provide platform for massive integration. |
| [54] | RIS based downlink multi user system. | - | Energy efficient design for transmitted power allocation and phase shift of the surface reflecting element. | E-maximization algorithms for base station transmit power allocation and RIS reflector values are designed. |
| [55] | Tabu Search scheme | Concave-Convex Procedure Heuristic method | Ternary scheme design for optimization: Tabu search method is used to solve optimization problem. | Resource optimization problems formulated focusing on signal to interference plus noise ratio for legitimate link. Results obtained through Tabu Search have provided excellent secrecy performance with less computational complexity. |
| [56] | VLC | VLC technique | Utilizing Multiluminaria Transmitters | Coordinated beamforming scheme is proposed for downlink interference mitigation among co-existing VLC atto cells. Features of CB scheme like lower requirement on network in terms of back bone traffic, ease in practical deployment are studied. |
| [59] | Orbital Angular Momentum based vertocose Communication | Equal probability Mode modulation. | Enhancement of spectrum efficiency with less number of RF channels: Huffmann coding based adaptive mode modulation scheme is used with CSI(Channel state information) . | Spectrum efficiency is maximized by developing equal probability mode modulation scheme. Joint power and probability allocation policy are developed for OAM based verticose Communication. AMM(Adaptive mode modulation ) EMM(Equal probability mode modulation) features are compared. |
| [60] | OAM | Multiple uniform circular arrays and Dielectric lens antenna | Design of Rx antenna with reduced SNR: Design of antenna is proposed with provision to change the antenna shape and size along with enhancement of performance. | Receiver SNR reduction issues are sorted out with multiple uniform circular arrays and Dielectric lens antenna methods and Guassian Beams are generated. |
| [61] | OAM(Orbital Angular Momentum) | Radio vortex wireless communication: multiple-mode OAM signal generation/adaptation/ reception; Long distance radio vortex wireless communications: OAM beam converging | OAM has the potential to increase the spectrum efficiency enhancement for LOS transmission, ultra-reliability with different modes, and anti-jamming with new dimensions. | This paper presents OAM for ultra-dense wireless networks. |
| [63] | 1. SBA(Service Based architecture)<br>2.SBA-RAN<br>3.End-to-End Mandate-Driven Architecture (MDA)<br>4. new 6G IP architecture | 1.Network slicing<br>2.Coexistence and interworking of multiple physical layer options as well as an independent evolution for each of them<br>4.Edge-based infrastructure for lower latency, software-defined anything (SDx) and additional computational and storage capabilities, AI, ML | 1.On-demand service deployment.<br>2. Internet of everything<br>3. Human-machine interfaces | 6G service-based architecture, new IP architecture, network and data analytics services and resource and energy efficiency parameters are studied. |
| [66] | Modulation and/or coding scheme (MCS) in Cognitive HetNet | Deep reinforcement learning (DRL), sensing-based approach. | Spectrum Sharing, Improve transmission capacity | In this paper, a cognitive HetNet are studied and proposed an intelligent DRL-based MCS selection algorithm for the PR to learn the interference pattern from STs. |
| [67] | In-band full-duplex (IBFD) | IBFD, which requires the utilization of multiple SIC techniques for different applications | Military jammer, provide physical-layer security ,massive multiple-input, multiple-output (mMIMO) architectures to reduce overall system complexity | This inclusion of IBFD technology within an upcoming wireless standard will provide a real time mapping of spectrum occupancy. |
| [68] | Radio Spectrum Management | Blockchains | Dynamic spectrum sharing applications | Blockchain distributed ledger up to date. It will enable secure, robust exchange of data. |
| [71] | 1.Passive Backscatter Device<br>2.System model for CABC system under flat fading Channels | OFDM<br>ML Detection<br>SIC based detection | 1.Receiver Design for CABC flat fading channels: ML based detection algorithms are proposed to design receiver. | Green Internet of Things is a key feature which is analysed.<br>Receiver design to extract information from both RF source and ambient |
| | 3.Receiver design for CABC under flat fading channel | | | Backscatter device is proposed. |
| [72] | Radio Management | Non orthogonal Multiple Access | 1.Mobility Management<br>2.Service Provisioning Management | - |
| [74] | Realtime AI | Specialized Neural Processing Units (NPUs), DNN | Unmanned Mobility, Human-centric services DNNs are increasingly used in live, interactive services, such as web search, advertising, interactive speech, and real-time video | NPUs are used for interactive services (1) execution of DNN models with low latency, high throughput, and high efficiency, and (2) flexibility to accommodate evolving state-of-the-art models (e.g., RNNs, CNNs, MLPs) without costly silicon updates. |
| [75] | Intelligent Radio Learning | 1.Regression models<br>2.K-Nearest neighbor<br>3.Super vector Machines<br>4.Bayesian Approach | Energy Harvesting : Value iteration and Bayesian Maximization approach used for energy harvesting. | Reinforcement Learning is utilized to infer decision making by mobile users under unknown network conditions. |
| [76] | Multi-access Edge Computing (MEC) MEC pushes 4C to the edge of the network. | Edge Intelligence, Cloud Computing, Big data, Network Edges | IT-based services and cloud computing capabilities at the networks edges | MEC servers collaborate to satisfy users' demand |
| [77] | Vehicle-to-vehicle (V2V) communication | Federated Learning for ultra-reliable low-latency communication (URLLC) in vehicular networks. | Unmanned mobility, autonomous driving and intelligent transportation systems | This work highlights the constraints on URLLC and characterized using extreme value theory and modelled as the tail distribution of the network-wide queue lengths over a predefined threshold. Leveraging concepts of federated learning, a distributed learning mechanism is proposed where VUEs estimate the tail distribution locally with the assistance of a RSU. |
| [78] | Data shuffling, Difference-of-convex-functions (DC) algorithm, linear coding scheme | Mobile edge intelligence | Data shuffling problem in wireless distributed computing system to improve the communication efficiency | The work proposes co-channel communication model for the data shuffling problem in wireless distributed computing system to improve the communication efficiency |
| [79] | Nano Communication Networks in TeraHertz Band | Network Slicing, Virtualization (NFV) and IoT technologies SDN, IoT, Fog computing | Nano-IoT | They proposed a scheme for communication between nano devices using terahertz band. |
| [80] | Molecular and nano–electromagnetic Communication. | Graphene based Nano Antennas | 1.Bandwidth and channel Capacity:<br>2.Information Modulation: Channel Sharing is done using asynchronous MAC protocols | THz band Communication Properties like path loss, Noise etc., are investigated. |
| [82] | | QAM,QPSK ,OFDM | 1.Recording High spectral efficiency.<br>2.Virtualization of end to end connectivity through mobile network to packet data network. | To achieve Very high bit rates geometric and probabilistic constellation methods are suggested. Intelligent traffic management and edge computing methods are suggested to maximize user welfare or quality of experience. |
| [86] | High altitude pseudo satellites(HAPS) for signal coverage | Overlapped and dynamic network topology. | 1.Signal coverage extension including non-terrestrial areas. Using THz communication band.<br>2.Virtual Massive MIMO-Single antenna usage is suggested. | The requirements such as high capacity /speed communication and extreme coverage extension, low power consumption and cost reduction and low latency, massive connectivity and sensitivity are suggested to go beyond 5G. |
| [91] | Analytical multi-ray channel modelling | Ray tracing | 1.Wide band channel capacity:<br>Decomposition of received signal in THz Band into sub-bands to compute the channel capacity by varying spectrum power.<br>2.Multipath propagation effects:<br>Various coefficients are evaluated using Fresnel's KED model to study the propagation effects on rough surfaces. | In depth analysis on the channel characteristics on the THz band. Reflection, scattering and Diffraction Coefficient are evaluated using KED model |
| [92] | Channel modelling in spatio temporal domain. | stochastic approach for indoor 300GHZ channel | - | The model facilitates system simulations under realistic channel conditions. |
| [93] | SISO,UM- MIMO channel modelling methods for THz communications beyond 5G | 1.Ray – Tracing<br>2.Finite Difference Time Domain | 1.Analysing Time varying Property: THz Channel needs to be developed by considering statistical movement of Transceiver ,propagation scenarios between LoS and Non-LoS | Analysis of large structures has been studied using Ray tracing and FDTD techniques. |
| [94] | 3GPP architecture | Ultra Dense Network Technology | 1.Resource Management<br>2.Mobility Management<br>3.Interfernce Management | Traditional network and UDN comparison done, efficiency of UDN to handle the traffic density is analysed. |

# II. Applications of 6G:

As shown in Table 2 and Figure 1, in comparison to 5G, the 6G technology will required to cater stringent requirements, such as higher capacity, wider spectrum band, higher power-efficiency, very low-latency control, three-dimensional integrated communications (3D-InteCom) to ensure the ubiquitous global network coverage, and connected intelligence through the capability of AI/ML [13, 16, 20]. In addition to the eMBB, uRLLC, and mMTC type services to be provisioned by 5G, 6G will require to support new services, such as Computation Oriented Communications (COC), Contextually Agile eMBB Communications (CAeC), Event Defined uRLLC (EDuRLLC), Big communications (BigCom), Unconventional data communications (UCDC), etc. [27, 41].

Further, 6G will also have to provision advanced services such holographic communication, personal monitoring, drone taxi, Internet of robots, wireless brain-computer interactions, AI/ML enabled autonomous systems, haptic communication, smart healthcare and biomedical communication, truly immersive XR, extended reality, digital replica, nano-IOT, bio-IOT etc. [16, 27, 41].

Table 2. 6G capabilities in comparison to 5G [7, 16].

| **Factors** | **6G** | **5G** |
|---|---|---|
| Peak data rate | > 100 Gb/s | 10 Gb/s |
| Individual Data Rate | > 10 Gb/s | 1 Gb/s |
| Traffic density | > 100Tb/s/km$^2$ | 10Tb/s/km$^2$ |
| Connection density | > 10million/km$^2$ | 1million/km$^2$ |
| Latency | < 1ms | 10 ms |
| Mobility | > 1000km/h | 350km/h |
| Spectrum efficiency | > 3x compared to 5G | 3~5x compared to 4G |
| Energy efficiency | > 10x compared to 5G | 1000x compared to 4G |
| Coverage percent | > 99% | Approx. 70% |
| Reliability | > 99.999% | Approx. 99.9% |
| Positioning precision level | Cm | M |
| Receiver sensitivity | < –130dBm | About –120dBm |
| Jitter | 1 µs | 10-100 µs |

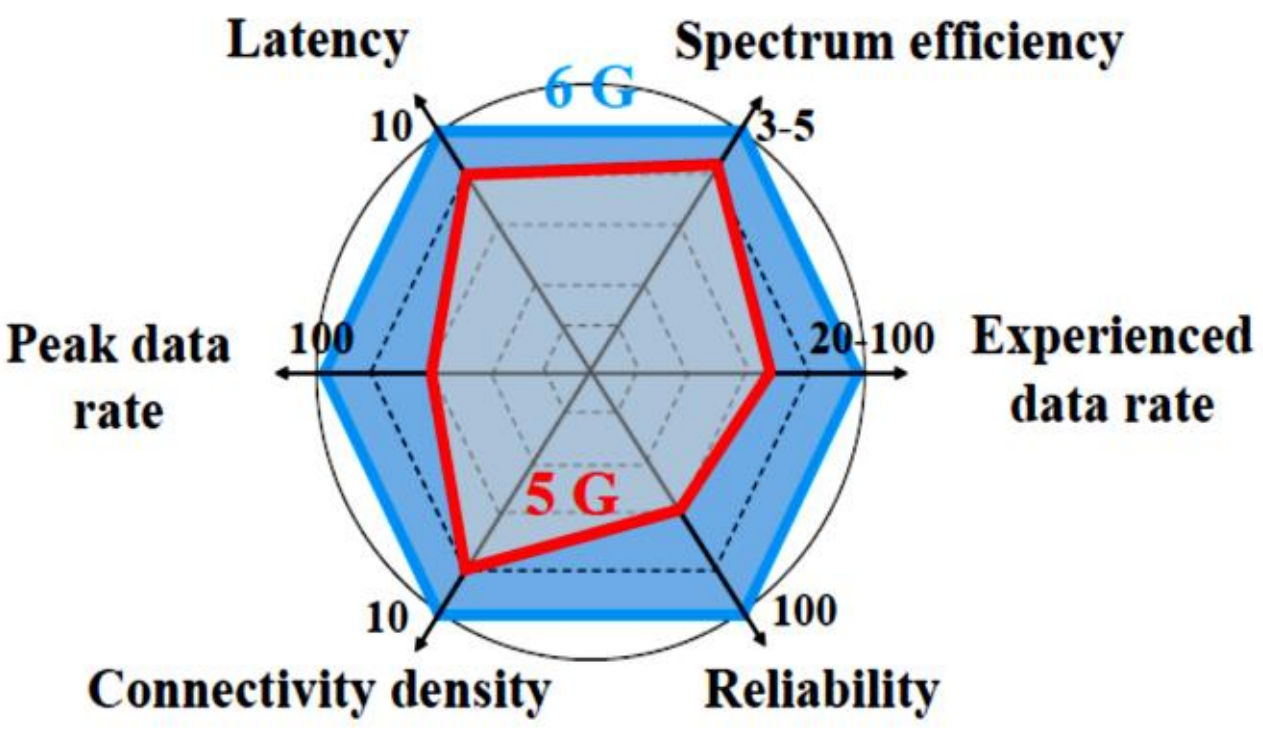

Figure 1. Comparison of 6G with existing 5G over various parameters [16].

# III. Focus Technological Trends

Among the numerous technological trends in the 6G technology, within this survey article's framework, we focus on the capacity enhancement techniques through efficient spectrum usage, which is summarized in the subsequent sections.

## a. THz Communications

The scarcity of spectrum and the limits in the present wireless communication systems' capacity can be alleviated by communicating in the THz band, which is the last unexplored span of the radio spectrum [42]. Specifically, the THz band dedicated to mobile communications is comprised mainly of the 275 GHz–3 THz band range, which is yet to be allocated for any services [43]. Further, from part of the THz band, one part between 275 GHz–300 GHz is defined for mm-wave, and the other part between 300 GHz–3 THz is defined as the far-infrared (IR) spectrum band. The latter band is a part of the optical band; however, it exists at the optical band's boundary and immediately after the RF band. Therefore, this band demonstrates similar characteristics as shown by the RF band [44]. The increased capacity envisioned for 6G communications can be accomplished by aggregating the THz band with the existing mm-wave band, which will result in approximately a 11x capacity increase [18]. Hence, the THz spectrum band includes massive amounts of available bandwidth and provides an opportunity to create extreme wideband frequency channels encompassing tens of GHz wide bandwidth. Thus, the ultra-high data rate communications required for 6G communications can be realized by resorting to the use of the THz spectrum, which has the potential to provide (i) tens of hundred GHz bandwidth resources, (ii) pico-second level symbol duration, (iii) integration of thousands of sub-mm-long antennas, and (iv) weak interference without full legacy regulation [16].

Communication in the THz band relies on advanced THz devices, of which, design of the THz transceiver is regarded as being critical to facilitate THZ communications [42]. In [55], the authors have already demonstrated an approach for generating THz frequency through a compact device that uses laughing gas to generate a THz laser with the frequency tuning ability. Traditionally, the THz spectrum's widespread use has been limited by the THz gap. However, currently, there exists two THz communication systems type viz., (i) **solid-state**: based on a frequency mixing technique, and (ii) **spatial modulation**: modulates the baseband signals directly over a continuous THz carrier signal. Further, various promising applications are envisaged, such as the Tbps WLAN (Tera-WiFi), Tbps Internet-of-Things (Tera-IoT), Tbps integrated access backhaul (Tera-IAB) wireless networks, and ultra-broadband enabled THz space communications (Tera-SpaceCom) [44]. Also, for the macro/micro-scale applications, the THz band can be employed for wireless connections in the nano-machine networks to enable wireless networks-on-chip communications (WiNoC) and the Internet of Nano-Things (IoNT). In this direction, global research activities are continuing, including the ICT-09-2017 cluster funded by Europe Horizon 2020, key programs funded by the Chinese Ministry of Science and Technology, and other NSF grants in the USA. The initial wireless communications standard, IEEE 802.15.3d (WPAN), which was published in 2017, operates at the 300 GHz frequency range supporting 100 Gbps and above data rate. Since the year 2019, the first and the second international workshops focusing on THz communications have been successfully held, while the momentum continues globally.

Another candidate for providing ultra-high bandwidth and capacity is the VLC, which enables short-range wireless communication and performs intensity modulation (IM) on light-emitting diodes or laser diodes as the transmitters and direct detection on the received light by photodetectors as the receivers [46]. VLC can achieve Gbps transmission through downlink using low-cost hardware, such as commercial light-emitting diodes, and a free, unlicensed spectrum [47]. Even though VLC (i) can only be used for the indoor scenarios owing to limited coverage range, (ii) requires an illumination source, and (iii) suffers from shot noise from other light sources [44], it is capable of integrating the space/air networks and underwater networks with the terrestrial networks to provide superior coverage [48]. Further, transmission through VLC (i) cannot penetrate opaque obstructions, (ii) does not generate electromagnetic radiation, (iii) is immune to external electromagnetic interference, and (iv) is cost-effective, as it utilizes illumination sources as the base stations [49, 50].

Overall, it is foreseen that the 6G systems will be based on a real cell-free architecture thus avoiding the handover issues and will offer seamless communication with improved quality of service and experience. Hence, a novel 6G architecture enables seamless interaction between numerous communication technologies, such as mm-wave communication, VLC, and THz communication. Figure 2 shows the electromagnetic spectrum and wavelength to be used for mm-wave communication, THz communication, and VLC, and Table 3 compares mm-wave communication, THz communication, and VLC over various parameters.

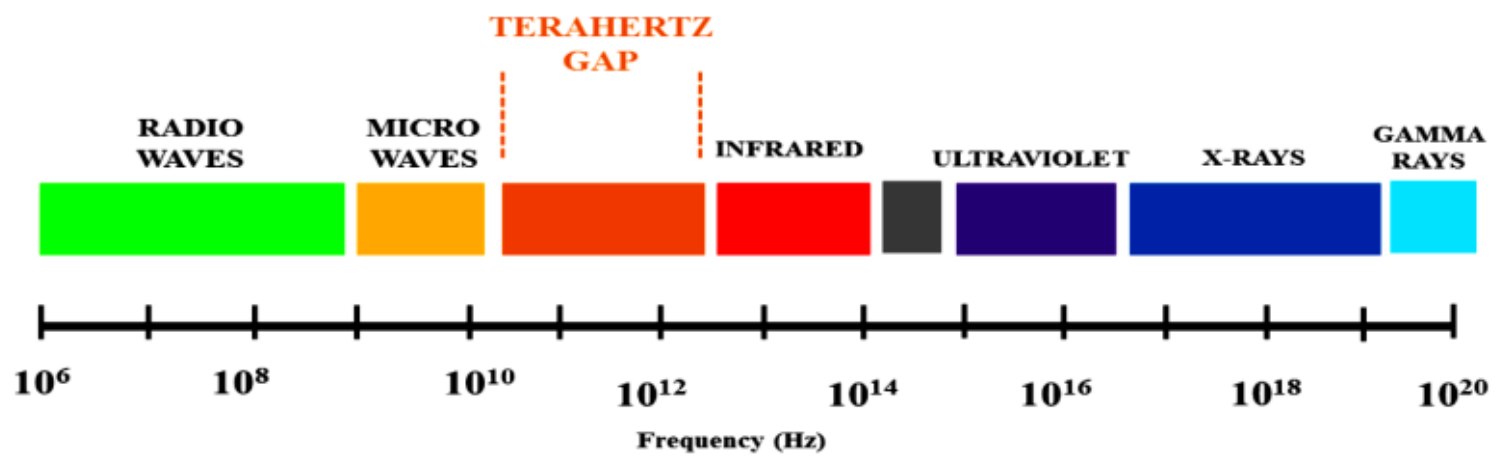

Figure 2. Electromagnetic spectrum and wavelength for mm-wave communication, THz communication, and VLC [39].

Table 3. Comparison of mm-wave communication, THz communication, and VLC [7, 10, 11].

| Parameter | Mm wave | THz | VLC |
|---|---|---|---|
| Available Bandwidth | MHz to several GHz | 10-100 GHz | 100 GHz |
| Transmission Distance | 3.8Km (2.36 miles) | No line of sight | Line of sight |
| Electromagnetic Radiation | Yes | Yes | No |
| Achievable capacity (data rate) | Utmost 10 Gbps | 100 Gbps | 10 Gbps |
| Spectrum Regulatory | Licensed | Licensed | Unlicensed |
| Penetration ability | Fog, Rain, Cloud, Skin | Special Opaque Materials | Transparent Materials |
| Inter-Cell Interference | High | High | Very Low |
| Cost | High | Expensive | Cheap |
| Transmission Power | High | High | Low |
| Diffuse Reflection Losses | Licensed | Licensed | Unlicensed |

Overall, THz communications and VLC are identified as the two most promising technologies for 6G communications due to their apparent advantages; however, many technical challenges still require solutions before their practical deployment.

Firstly, the existing challenges for THz communication are as follows:

1. Regarding the hardware components, it is difficult to design efficient RF circuits (e.g., mixers, oscillators, amplifiers, etc.) and high-performance modulators that will demonstrate efficient operation at the THz spectrum [7]. Current research has identified graphene as a potential material to be used for hardware design as it has high thermal and electrical conductivities and plasmonic effects [51]. Even though research on the development of chip-scale THz technology has picked up pace in the past decade, much research is still required on solid-state electronics for THz band [52]. Also, low power consumption and high Giga-samples/sec enabled analog to digital and digital to analog converters, which can transfer data at Tbps with low power consumption, need to be designed for THz communication [16]. Further, THz communication will require a higher number of antennas compared to mm-wave communication, which will pose many significant practical problems.

2. THz Reconfigurable Intelligent Surfaces (RIS) have been proposed lately as an effective solution to overcome the interruption of obstacles by steering waves adaptively in the customized directions. In addition to mitigating the LoS blockage issue, the RIS can improve coverage by (i) acting as relays and security in THz systems, and (ii) directing links to surpass eavesdroppers. However, much research still needs to be conducted in regard to RIS for THz communication. Lastly, for supporting GHz wideband channels, orthogonal frequency division multiplexing (OFDM) is a favoured technique; however, with the aforementioned limitations in the hardware for 6G, research must be directed towards exploring alternate methods.

3. Implementation of accurate channel models for both, indoor and outdoor situations is required for efficient THz communications. However, the design of such a channel model is challenging since, in comparison to the lower-frequency channels, the THz channel will be severely affected by (i) highly frequency-selective path loss, consisting of spreading loss and atmospheric loss (or molecular absorption loss), which occurs due to the presence of absorption lines of oxygen and water existing in the THz frequency range, (ii) diffuse scattering and specular reflection, (iii) diffraction, human shadowing, and line of sight probability, (iv) weather influences and scintillation effects, and (v) mutual coupling and near-field effects [16]. Further, for efficient channel estimation, ultra-massive arrays of the antenna will be required, which will pose the challenges of (i) spatial non-stationarity, and (ii) design of the feed network and antenna elements required to support the wide bandwidth [7]. Also, on the one hand, the use of such ultra-massive arrays will result in precisely focused beams; on the other hand, it will lead to communication which will be subject to LoS and precisely reflected paths rather than scattering and diffracting paths. Hence, it will be required to optimize the architecture to be used for beamforming in view of provisioning high dynamic-range and flexibility at a reasonable cost and energy consumption [16]. Although the research community has identified metamaterial-based Antennas and RIS as the beamforming architectures [53, 54]; much research is still desired in this regard. The survey in [41] has reported that 6G will result in the use of the Rayleigh channel model rather than the Rician channel model, and the survey in [50] suggests that a mixed model may be the solution to the THz channel modelling, i.e., rather than reusing the narrowband fading models, such as Rayleigh or Rician, a statistical impulse response model needs to be developed in the THz band. Also, high accuracy with high time and resource consumption can be obtained for a deterministic channel model, whereas; a statistical channel model demonstrates low computational complexity with a lack of accuracy. Hence, there is a need to develop a hybrid channel model for THz communication by strategically combining the benefits from two or more individual approaches.

4. To accomplish the directional networking in THz communication, directional antennas will be favored due to their capability to concentrate the THz signal energy in a specific direction. Directional antennas support long-distance communication and minimize neighboring node(s) interference. Further, Ultra Massive-MIMO can also be implemented to enable spatial multiplexing for increasing the capacity, in conjunction with beamforming. However, many challenges exist for THz band directional networking, including efficient strategies for neighbor discovery for a time asynchronous system, algorithms for topology control including the optimized trade-off between the node degree and the jump stretch, high access capacity enabling and low resource utilization ensuring multiple access control protocol, etc. [7]. To achieve efficient Ultra Massive-MIMO communications, spectral-efficiency, and energy-efficiency, hardware complexity needs to be considered jointly. Besides, the issue of sparsity also needs to be rigorously treated in view of exploiting the spatial multiplexing design considering the possible spherical propagation of THz signals over the extensively spaced antenna arrays.

5. Mesh networking will be essential to extend the communication reaches and network coverage for THz communications [44]. However, owing to the THz links being prone to breakdown, the challenge lies in delivering the stable end-to-end networking across the multi-hop paths. For ensuring stability in mesh networking, the critical step will be to proactively conduct the link and the topology management via an efficient control signalling method either through in-band, dual-band signalling operations, or both. Hence, in THz mesh networking, the (i) cross-layer design between the link layer and the physical layer will be necessary to ensure the quality of service requirements at both, the link level and the end to end level, and (ii) simplifying the interactions across the protocol stacks will assume importance owing to the ultra-high data rates and low latency requirements.

6. The total traffic flowing via the 6G network will increase once there is a transformation from multi-Gbps capacity to the THz capacity. This will in turn lead to a heavy burden within the transport layer over the congestion control and reliable end-to-end transport. Hence, there is a need to revise the existing TCP congestion control window significantly in view of handling the dynamics of the traffic. One approach towards this end is merging the network and transport layer with the data link layer. However, this will mandate the development of novel routing and transport protocols, which will improve the protocol stack efficiency and will also enable the joint operation and optimizations of medium access control, routing and transport functions.

7. Concerning the Medium Access Control (MAC), there exist major issues, such as (i) increased problem of aligning a transmitter and a receiver owing to the very focused and directed beams, (ii) transmission coverage and network discovery problem due to the control channel having to search the frequency band for communication (i.e., mm-band or THz) and the modes of the antenna which are full or semi- or omni-directed, (iii) no specified boundary of the cell, and (iv) monitoring of interference and simultaneous transmission scheduling with spatial reuse for dense networks [39]. The aforementioned issues can be mitigated by implementing the long-user-central window (LUCW) method simultaneously with successive interference cancellation at a receiver terminal. The aforementioned, together with beamforming technologies, forms the THz non-orthogonal multiple access (NOMA), which has excellent potential to optimize the spatial, spectral, and power resources; however, it requires further investigation.

8. In regard to the coding techniques, the existing forward error correction (FEC) codes need to be enhanced so that the channel errors in THz communication can be overcome [39]. The enhanced FEC codes for THz communication also need to be validated and demonstrated in the advanced THz chipset. Once the error sources in THz communication have been comprehended, advanced state-of-art codes such as turbo, low-density parity-check, and polar codes can also be evaluated as possible candidates. As an alternative to existing coding schemes, disruptive types of ultra-low-complexity channel coding schemes can also be developed for THz communications. In regard to designing effective error control policies, it will be mandatory to characterize the nature of channel errors, which are mainly affected by stochastic noise, interference, and multi-path fading.

Secondly, the existing challenges for VLC communication are as follows:

1. To enhance the bandwidth, the use of commercial light-emitting diodes limits the modulation bandwidth and demonstrates a slow modulation response. This, in turn, limits the capacity which is achievable by the VLC based systems [47]. Therefore, it is required to develop the advanced pre- and post-equalization methods to provide solutions to the modulation bandwidth limitation of VLC enabled systems.

2. For high-speed VLC enabled systems, OFDM is considered as a powerful modulation technique [48]. However, since the current via a light-emitting diode and the light emitted through a light-emitting diode bears a non-linear relationship, the limited dynamic range and the high peak-to-average-power ratio of a light-emitting diode emerges as the main challenge for an OFDM enabled based VLC system [47]. Existing studies have demonstrated that the advanced pre- and post-distortion techniques included with OFDM could be an efficient solution to countermine the nonlinearities effect.

3. The data rate for VLC enabled systems can be increased by implementing the MIMO techniques. However, it is arduous to realize MIMO based systems for VLC [55] since (i) the transmitter and receiver have the same paths, which limits the diversity gains, and (ii) receivers enabled by MIMO VLC are challenging to realize [56]. Hence, techniques such as, multiple user pre-coding and aligned beamforming to reduce the interference in MIMO VLC need to be further investigated.

The study in [41] has also envisioned a system implementing a hybrid VLC, and the THz communication technique system ushers in a sturdy communication technology sustaining the ambient light, thereby eventually reducing the signal-to-noise ratio of the VLC system.

## b. Modulation Formats

It must be highlighted that the THz spectrum will encounter a distance related bandwidth owing to which appropriate modulation formats need to be developed according to the application as per the desired length of transmission. Modulation formats, such as pulse amplitude modulation (PAM), pulse position modulation (PPM), and on-off keying (OOK), can be explored for the short-distance communication. These formats exchange approximately 100 F-sec long pulses which can be generated and detected easily using photonic and plasma wave devices. In the case of long-distance communication, there can be a dynamic adaptation of the waveform to be transmitted within the THz spectral window which will lead to the desired distance adaptive multi-wideband pulse modulation. Further, due to the large bandwidth availability, the spread spectrum modulations can also be explored. During the proposal of 5G, the implementation of non-orthogonal multiple access was put forth [2]; however, research has shown that, to obtain the desired high spectrum-efficiency and enhanced capacity in 6G, new modulation formats will be required [7, 16]. In this regard, existing studies have identified that, in addition to comprising of linear momentum, any electromagnetic signal is also constituted of angular momentum [57]. This implies that for light propagating through space both, the electric and the magnetic fields have individual orthogonal axes that give rise to two rotation types, namely, the spin angular momentum (SAM) and the orbital angular momentum (OAM) [58]. Specifically, OAM is a wavefront that demonstrates the helical phase and consists of multiple topological charges known as the OAM modes [59]. The authors in [7] have shown that the (i) OAM waves are centrally hollow except the OAM mode 0, which corresponds to the traditional electromagnetic wave, and (ii) central hollow pattern and divergence degree increases with an increase in the order of the OAM mode. Further, experimental research has demonstrated that for the electromagnetic signals, through the use of a transmit antenna array, it is possible to generate various modes of OAM concurrently, and multiple data streams can then be modulated on different OAM modes [60]. Further, numerous OAM modes are orthogonal to one another, and multiplexing and demultiplexing them together has been identified as a novel method to enhance the capacity as desired in 6G communications [61]. Lastly, the study in [58] has demonstrated that OAM is a special case of spatial multiplexing in regard to capacity and antenna occupation, and hence, through OAM, high order spatial multiplexing can be accomplished which is impossible with conventional MIMO systems.

Overall, OAM is a prime candidate to achieve the desired high capacity for 6G communications; however, the following major technical challenges still require solutions before OAM is practically implemented:

1. The OAM beams demonstrate divergence [57] due to which the 6G systems enabled by OAM will show limitations in regard to reception efficiency and signal propagation for long distances. However, to ensure communication efficiency, the OAM beams' convergence will be desirable for which techniques, such as antenna-based anti-divergence, have been proposed [61]. However, as the divergence degree reduces, even after convergence, OAM beams have been shown to have a central hollow structure after relatively long-distance transmission, and research needs to be directed towards finding solutions for the same.

2. Existing studies have shown that OAM can be implemented for the LoS scenarios, such as indoor communication requiring massive data [58]. However, for the non-LoS scenarios which result in fading, a variation of the OAM modes wavefront with the OAM waves refraction and reflection results in increased OAM communication model complexity [60]. Therefore, accurate models for OAM communication in non-LoS scenarios need intensive investigation.

3. The OAM waves are phase-sensitive, owing to which existing research has proposed the implementation of methods that detect the current [59]. However, such methods assume aligned transmitters and receivers, whereas, in practicality, there is the misalignment of the transmitter and the receiver, which poses immense challenges in identifying the various OAM-modes. Although methods such as, phase turbulence compensation have been proposed for OAM based transmission in misalignment scenarios [16], much research is still desired to find efficient solutions.

## c. Spectrum sharing and efficient Radio Access Technologies

The foundation of any communication technology is a spectrum. The past few decades testified an enormous increase in the traffic volume and the corresponding data rate, which has resulted in the requirement of large amounts of spectrum resources. As one of the foremost objectives of 6G communication is to provision Tbps capacity, it will be mandatory to operate at very high frequencies to make large bandwidth available to the users. Further, the spectrum resources are critically essential to guarantee seamless communication coverage; however, these resources remain scarce in every generation. Further, often, the exclusive service providers or licensees do not actively use the allocated spectrum, which results in the underutilization of the precious spectrum resources. In such cases, permitting other parties (service providers) to utilize the unused spectrum allows for efficient use of the constrained spectrum resources. It is also often observed that the institutions responsible for framing the spectrum regulations consider a deviation from the traditional exclusive spectrum licensing method in view of obtaining an improved use of the scarce spectrum resources. One such method is that of spectrum sharing which allows multiple users to utilize the available spectrum [7].

Till the 5G era, complete occupation of the spectrum was ensured by the implementation of dedicated spectrum allocation, which, however, resulted in a low rate of utilization [62]. For 6G communications, higher spectrum-efficiency will be a significant requirement that will demand much efficient spectrum management techniques. One such method is Cognitive Radio, in which, by efficiently sensing the spectrum and managing the interference, the same spectrum can be shared by multiple users. Specifically, in view of enhancing spectral-efficiency, Cognitive Radio allows using the less used or unlicensed band by the secondary users [63]. The efficiency of a Cognitive Radio system can be further enhanced by using symbiotic radio has also emerged as a promising technique in which spectrum resources can be shared efficiently between multiple heterogeneous user systems via intelligent cooperation [62]. A symbiotic radio may comprise various symbiotic mechanisms such as parasitism, mutualism, and commensalism. The study in [64] on the Symbiotic Radio network is based on ambient backscatter communications in which the network devices use ambient radio frequency signals for transmitting the information without the need for any active radio frequency transmission in battery-free communication.

However, even with such advanced techniques and their implementation; there still remain the following major challenges in regard to spectrum sharing:

1. The high correlation of various operators' networks limits the spectrum's opportunity to be shared. Over short periods, the networks of multiple operators demonstrate varied data traffic patterns and in such scenarios, sharing of the spectrum can be more effective if dynamic spectrum sharing techniques are implemented compared to static or semi-static methods. However, the major challenge will be to avoid spectrum utilization collision due to various users simultaneously permitting the spectrum's dynamic access [7]. To enable the aforementioned, complete network state and spectrum access information will be required, which, in a real-time scenario, will result in excessive overhead [65]. Hence, existing spectrum sharing algorithms do not efficiently adapt to a real-time scenario. The AI-enabled studies have demonstrated that to prevent collisions, spectrum utilization can be predicted with limited information exchange [27]. Deep reinforcement learning-based dynamic spectrum sharing algorithms have been proposed to learn the primary systems' hidden patterns [66]. However, to ultimately ensure the real-time applicability with complex spectrum sharing scenarios, much research needs to be directed towards advancing the soft computing overhead AI-enabled algorithms.

2. Many other spectrum sharing methods can also be implemented in 6G communications. These could include full-duplex communication, non-orthogonal multiple access, and spectrum sharing in the unlicensed spectrum. Among these methods, the free-duplex method has been identified as an efficient technique to minimize the existing imbalanced use of spectrum by providing complete freedom degree. Allowing the sharing of spectrum resources between the transmitter and the receiver in space, time, and frequency between transceiver and receiver links [67]. The report in [16] has also identified that deviating from the principle of mutual exclusiveness (i.e., restriction of downlink and uplink to use mutually exclusive time-frequency resources) would enable the dynamic adaptation of the duplex technology. Hence, much research is still solicited in the advancement of the duplex technology.

3. The 6G applications will result in massive connections for which efficient techniques need to be formulated, which will focus on reducing interference and enhancing the system's performance. In this regard, blockchain and deep-learning technologies have been demonstrated as the efficient techniques [68]. Specifically, blockchain provides multiple features such as distributed network structure, consensus mechanism, etc., which are not present in the existing structures [69]. It also produces a secured and provable approach for spectrum management by verifying transactions, preventing unauthorized access, etc. Having features such as, decentralized tamper-resistance and secrecy, makes blockchain an ideal candidate for multiple applications in 6G communications [70]. In 5G, the main limitation of blockchain concerns the throughput [2]; however, with the use of consensus algorithms, novel architecture, and sharing methods, and an increase in the network block size, the limitations can be minimized in 6G communications, which paves the way to further research in this direction. Hence, advanced frameworks and mechanisms for incorporating intelligence within spectrum sharing need to be developed.

4. The emergence of symbiotic radio poses many interesting questions, as a frontrunner technology to improve system performance through intelligent inter-subsystem cooperation. Even though at an infant research stage, investigation of every symbiotic mechanism is required to find the theoretical limits of fundamental information and transmission theory with interference management [71]. Further, for sizeable symbiotic radio networks, wireless access and multi-dimensional resource allocation are also challenging aspects that need further investigation through AI techniques.

## d. Intelligent Networks

For 6G communications, the final goal is to automate the network evolution, which is possible via the intelligent networks. Regarding the services, there is a need to transform the network-as-a-service to the network-as-an-intelligent service, enabling 6G systems/networks to adapt the various parameters dynamically in turn offering enhanced performance. Further, the advantages of introducing intelligence within the 6G functions will provide operational, environmental, and service intelligence [41]. Initial intelligence requires a relatively isolated network entity with an intelligent capability to adapt the configuration according to multiple a-priori options in a varied but static manner capable of solving unexpected scenarios [72]. With network evolution resulting in a complex and heterogeneous system, new AI techniques will be mandated, which have the properties of self-awareness, self-adaptiveness, and self-interpretiveness [30]. This evolution, in turn, will require the existence of AI techniques within the entire network so that all the network components have autonomous connect and control for adapting and solving unexpected scenarios. Specifically, the envisioned future requirements of 6G, which encompass edge intelligence, will result in the evolution of edge computing towards an AI platform capable of offering intelligent services to be delivered to the edge devices over the static or dynamic network [27]. This will require the deployment of hardware to implement edge ML/AI algorithms, compared to current implementation of static and centralized AI/ML algorithms over the currently deployed cloud computing hardware [13].

AI technologies can also enhance wireless communication in 6G. In the current-generation wireless communication, the physical layer suffers significantly from multiple channel impairments such as, fading, interference, etc., and many hardware impairments such as, amplifier distortion, local oscillator leakage, etc. To communicate reliably and efficiently in the presence of combined impairments, it is required to control and optimize multiple design parameters jointly. Further, the complexity of wireless communication systems decreases the practical implementation of the end-to-end optimization. Hence, current systems split the complete chain into many independent blocks wherein, every block exists as a simplified model; however, such a technique does not accurately capture the practical systems' features. In view of the aforementioned, advanced AI techniques make it possible to conduct the end-to-end optimization of the physical layer's complete chain, from the transmitter to the receiver. In 6G communications, the physical layer will be intelligent so that the end-to-end system will be capable of self-learning and optimization, combining AI technologies, advanced sensing and data collection, and domain-specific signal processing methods. Existing studies have shown that deep neural networking can be used to train the transmitter, channel, and receiver as an auto-encoder such that, the transmitter and the receiver can be simultaneously optimized.

Overall, in 6G and the enabling smart applications, AI techniques are also required to provide intelligent MAC protocols and intelligent transceivers, thereby making AI an integral part of the 6G wireless network. Further, the intelligence demanded by 6G communications can be enabled by the following methods.

i. As already highlighted, the current static feed based AI techniques over centralized cloud networks will not be capable of providing low-latency and high intelligence-enabled 6G networking. In response to the aforementioned, the research community has already begun the collaborative work, via research labs, of developing intelligence at the edge via the AI techniques, which can predict, infer and decide based on real-time feed [73]. In such advanced AI techniques, the following occurs: (i) data is trained unevenly and is distributed over large amounts of edge devices, (ii) every edge device is given the access to a fraction of the data and a limited computation and storage power such that (a) these edge devices can exchange their locally trained models as opposed to exchanging their private data, and (b) collective training and inference is conducted, and iii) to avoid the storage of massive data which has been monitored, data abstraction, cleansing, and dimensionality reduction of network data assumes vitality. However, 'AI at the Edge' is still in the nascent stage of research and requires much investigation concerning secure and reliable communications and on-device resource constraints [13]. Lastly, the industry has also commenced the designing of specialized real-time AI processors [74].

ii. From a physical layer perspective, there is currently only initial intelligence wherein, the devices and the transceiver algorithms are designed together. The latest emergence of advanced radio-frequency and circuit systems hardware is on course to drive 6G communications towards tracking and completely exploiting the hardware's tremendous upgrade at the device level and the base station level. This will necessitate a split of the currently merged algorithm-hardware architecture [39]. With the advent of Intelligent Radio, the hardware and the algorithms applied to the transceivers can be segregated such that the transceiver algorithm configures according to the hardware capability via the ability to automatically evaluate the capability of the transceiver over which the protocol operates [40]. Specifically, the transceiver algorithm will be the software which will execute over an operating system to estimate the hardware's capabilities and automatically estimate the corresponding analog parameters. Further, the operating system will then use an interfacing language to configure the transceiver algorithms based on the hardware and the AI techniques. From the physical layer point of view, Intelligent Radio has the ability to (i) access the available spectrum resources, (ii) control the transmission power, and (iii) adapt the transmission protocols via AI techniques [75]. Further, compared to the existing cognitive radio technology, which is based on conventional modulation or coding methods, Intelligent Radio will involve the use of Deep Neural Networks (DNNs) for an intelligent adaptation to the environment and the hardware. Specifically, initially, the DNN algorithm will be trained at both, the transmitter and the receiver end by sending labelled training data. Then, the information bits will be transmitted upon the fulfilment of the desired performance. Overall, the application of Intelligent Radio in 6G has been envisioned to reduce the implementation time and the cost of new algorithms and hardware [40].

iii. For the requirement of real-time interaction in 6G, centralized AI/ML-based on one-time training can be implemented; however, the model trained through such algorithms does not produce good results owing to the frequent addition of new data. In contrast to the current centralized training and inference conducted over the cloud networks, 6G will necessitate a mostly decentralized system in which intelligent decisions will have to be arrived at over multiple granularity levels. To enable the aforementioned, by aptly dividing the model and the data, the advanced distributed AI techniques provision the network resources via simultaneous training [76]. The research community has already begun work over distributed AI techniques by developing federated learning in which the model is of the shared global type wherein, training of the model is conducted at the edge considering patterns of the local sample. These are further communicated over the centralized cloud network to conduct the averaging of the model, enhancing security and privacy [77]. Federated learning grants the benefit of considering newly added data training; however, it suffers from fairness issues that can be resolved by implementing the q-fair federated learning [77]. Also, for federated learning, a substantial effort is required to develop efficient methods for distributed optimization between large numbers of devices that share a standard model to be trained via distributed data across a large number of interconnected devices and by utilizing specific models. Further, for implementation in 6G, the research community has also identified ML techniques such as, quantum ML which combines quantum physics and ML to enable fast training of the ML models, and Meta-learning, which enables ML models to learn; however, incurring complexity in the design due to the various ML models having varied nature. Therefore, 6G will address the communication challenges for large-scale distributed ML, communication efficient distributed training, communication efficient distributed inference, etc. [7, 16]. In this regard, the uplink and downlink communication strategy's joint optimization was developed for shuffling the locally computed intermediate values across mobile devices [78].

iv. In regard to the aforementioned, AI techniques are a trump card for 6G, which can modify the current network architecture into the required dynamic ubiquitous network providing extended coverage, supporting THz communication, converging the computing, varied provisioning needs of massive applications and improving the overall performance. Further, with the inclusion of advanced AI techniques, 6G will be able to interconnect massive amounts of devices from diverse environments such as, terrestrial, submarine, and space, to provide them an unprecedented quality of service. Although there are aforementioned significant benefits of the AI technology, there still exist the following challenges before the practical deployment:

i. There exist the pros and cons of centralized cloud-based AI and edge AI. For example, with computing and AI capabilities at the cloud, on one hand, the resource-limited end-user devices have the benefit of rich computing power and storage in the cloud and edge. On the other hand, the backbone networks incur heavy load, resulting in unpredictable transmission latency. However, if the computing and AI capabilities are deployed on the edge, load on the backbone networks and delay in transmission can be significantly minimized. However, end-user devices will incur massive computation delays owing to limited computing power and storage on edge. Therefore, the question to be answered is whether the computing and the AI capabilities are to be included in the cloud or around the edge? In this regard, collaborative intelligence and cloud edge computing needs in-depth investigation.

ii. The tremendously increasing computing power of the edge devices will dictate the effectiveness of the AI techniques. Specifically, in 6G, on the one hand, radio communication will shift to the THz band, which will result in tremendous cost and higher power consumption of the hardware elements, in turn significantly affecting the transceiver design and algorithms; whereas, with the pervasive nature of devices to be used in 6G, there will be a limit on the (i) resources for communication, (ii) computing power, and (iii) storage which is sufficient for the IoE application. Hence, to minimize the performance degradation, advanced AI techniques need to be developed, which will enable the holistic design of communication, sensing, and inference. In this regard, firstly, there is a need to research on novel edge AI techniques for 6G devices. Secondly, for the design of a complex system/network such as 6G, the hardware and the algorithms (software) must collaborate effectively. Thus, research needs to be directed towards the hardware algorithms co-design to develop the hardware-efficient transceiver structures that also function effectively with the algorithms via AI techniques.

iii. The high heterogeneity of 6G concerning the type of device, spectrum occupied, kind of infrastructure and service, etc., will prohibit intelligent coordination of communication, data caching, and computing resources. Further, for ubiquitous intelligence within the entire system, the network architecture is desired to be flexible and straightforward through software definition and reconfiguration. In this regard, dynamic convergence needs further investigation by designing a software-defined and re-configured radio access network architecture resulting in the dynamic re-organization of communication, data caching, and computing. Overall, in addition to embedding intelligence across the entire network, it will also be required to embed the logic of AI within the network structure such that the perception and inference interact systematically. This enables all the network components to connect and control autonomously to recognize, adapt to, and provide unexpected scenarios. Finally, the expectation from intelligent networks is to ensure the autonomous evolution of the networks.

iv. Recently, new networking technologies such as nano-networking, bio-networking, and 3D networking have been identified as the prominent enablers for intelligent networking in 6G [7, 79, 80]. Although these technologies will be integral to 6G networking, there are multiple challenges regarding their implementation. Compared to traditional networking, both nano-networking and bio-networking have significantly different properties that require novel physical layer and routing methods to be formulated.

Further, these technologies will also need the development of efficient nano-devices and bio-devices. In comparison to 2D networking, for 3D networking, new models must be developed since the core network will comprise of terrestrial, submarine, and satellite system, which will be a novel ubiquitous model.

## e. Role of ML to accomplish Enhanced Spectrum-Efficiency

As stated previously and also delineated in the International Mobile Telecommunications (IMT)-2030, the requirements of 6G's unprecedented outcomes are ultra-high data rates (peak data rates of 1 Tbps and client endured data rates of 1Gbps) and higher network capacity, ameliorated spectrum-efficiency (5x), and enhanced energy-efficiency (100x). Moreover, as particularized in IMT-2030, 6G will also accomplish enhanced Quality of Service and Experience (QoSE) and system throughput (99.999%), ultra-high user connectivity density (10^7 devices/km^2), large area traffic capacity (Gb/s/m^2), ultra-low latency (0.001-0.1 ms), high-mobility (≥1000 km/h), and reliability (99.999%). Therefore, compared to 5G, 6G's network complexity will be very high. Further, the applications and features of 6G such as, AR/VR, Internet of Nano Things (IoNT), Tactile Internet and Haptic Communication, Deep Connectivity, Holographic and Ubiquitous Connectivity, Smart healthcare, IoE, Automation and manufacturing, Super-smart society, connected robotics, autonomous systems, Wireless brain-computer interactions, five sense information transfer, Unmanned Aerial Vehicle (UAV), etc., will further increase the network complexity. Also, such a dense and complex network will be difficult to maintain manually. Furthermore, the problems such as effective usage of the physical layer, efficient MAC and resource allocation techniques, network routing protocols, network security issues, and application-oriented automation will be highly challenging to address. Therefore, the conventional network management will not work with 6G communications, and it is essential to achieve the zero-touch architecture using ML algorithms. Moreover, the current generation of wireless networks employs complex mathematical models, resulting in higher computational complexity and reducing accuracy. On the contrary, the revolutionary application of ML in 6G replaces the complex mathematical models with simplified self-organizing mathematical models. This reduces the computational complexity and improves accuracy, energy, and spectrum-efficiency. Also, there is an open challenge to obtain the initial training data set for 6G, which is currently unavailable. In this regard, one proposal is that with the commercialization of 5G, there will be broader set of data available which can be extrapolated for 6G communications [27].

ML is a sub-branch of AI which has the cognitive capability to perform computational task and predict the outcomes based on the predefined ML algorithms. The ML algorithms efficiently predict the results found by the learned ML agent during the training phase. The accuracy depends on the computational capability of the machine and the available training datasets. Figure 3 summarises the ML paradigm into four subcategories viz., (i) Supervised, (ii) Unsupervised, (iii) semi-supervised or reinforcement, and (iv) deep learning. In supervised ML, the ML model is trained by the inputs' given labelled data set and their known mapped outputs. Hence, in the case when a new input arrives, that input will be mapped to the appropriate output. On the other hand, in unsupervised ML, the training data sets' labelling is not conducted. ML model learns and classifies the input samples into the associated clusters, and based on the clustering; it predicts the outcomes. At the same time, in reinforcement learning, which is also known as semi-supervised ML, the ML agent interacts with the reward function, mapping the inputs to the output based on the highest reward. Furthermore, deep learning is a subcategory in ML in which the model can have multiple hidden layers within the system, hence predicting the outcomes efficiently using a rewards function and improving the prediction accuracy [43]. Furthermore, deep learning can be divided as RNN, multilayer perceptron's support vector machine (SVM), Gaussian Mixture Model (GMM), etc. which can be used in wireless communication [24, 72]. Lastly, there exist many other models which are the candidates for 6G, and these are described next, and are also summarized in Figure 3.

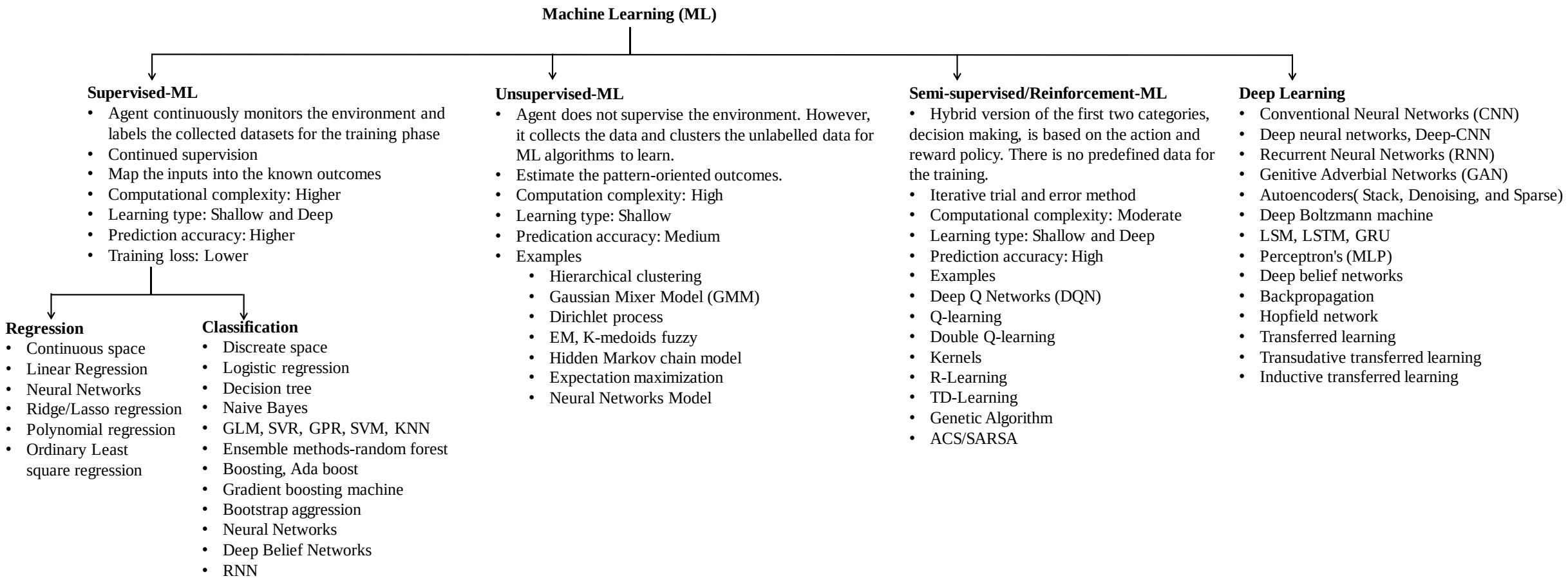

Figure 3. Classification of the state of the art Machine Learning techniques.

## Bayes' probabilistic model in ML

In recent times, the Bayes' probabilistic model has attracted the attention of many researchers to focus on the applications in wireless communications. However, the computational complexity is very high. The vibrational encoders, Bayesian Gaussian process, Monte Carlo, and Markov chain models are the popular models in this category.

## Kernels

As mentioned earlier, the requirements of 6G will lead to higher interference and will also impose the performance bottleneck. 6G is also expected to suffer higher path loss, Mie scattering, human shadowing impacts, hardware mismatches, and other impairments in wireless communication. Therefore, kernel Hilbert space looks a promising solution to mitigate the impairments and demonstrate a lower approximation of error considering the large interference non-Gaussian environment. Therefore, this solution is proposed to be used in 6G communications and applications such as, resource allocation, detection, tracking, estimation, localization, and is also expected to be used along with the other methods such as Monte Carlo and deep learning methods to achieve.

## Federated/Transferred Learning

In the conventional ML learning method, wireless communication devices are expected to transfer the collected data for learning, increasing the network's traffic load, and delay. Therefore, it is not suitable for the dynamic applications of 6G, targeting the high speed and low latency. Whereas, in transferred or distributed learning, the local devices are collecting the data, training the local ML agent, and sharing the trained state to the global server. Based on the local states, the server trains the global agent, and the global states will be transferred to all the communication devices. The significant advantage of the transferred learning is that the communication devices do not share the entire data set to the global server, thus improving the spectrum-efficiency and reducing the network traffic load.

In reinforcement learning, the ML agent interacts with the environment, and collects the required information to train the agent based on the action versus reward function, thus maximizing each resource allocation action's reward. Various versions of the neural networks such as AAN, CNN, or RNN can be used to combine reinforcement learning in mapping the function, such as RELU or sigmoid, thereby maximizing the reward for the action policy. Moreover, reinforcement learning along with neural network and deep learning can be used in multiple resource allocation problems such as, spectrum- and energy-efficiency, power allocation, adaptive modulation formats assignments, adaptive pencil beam steering, interference mitigation, and impairment mitigation for the THz communication.

**ML layered architecture for 6G**

Layer-0 (L0) mainly deals with the physical deployments of the devices channel performance and data forwarding over the 6G networks. The present generation wireless communication networks employ sophisticated models to determine channel performance. However, the complexity involved in such mathematical models is very high, and the outcomes are not accurate. Therefore, it is proposed to employ and investigate the ML techniques in 6G to overcome such issues. Such ML method can be employed in encoding, modulation, formatting, decoding, synchronization, positioning, and other functions related to L0 of wireless communication.

**Channel estimation and beamforming**

THz communication channels are impacted by a higher path loss, atmospheric absorption, rain attenuation, human shadowing effect, and building penetration losses. Therefore, accurate channel estimation is essential to identify the best channel performance, and the application of deep learning plays a vital role. Moreover, to mitigate the higher impairment impact on the THz channels, channel estimation and beamforming works closely to achieve higher performance. Establishing the adaptive pencil beams with high directivity extends range delivering higher SINR. This can be achieved using the deep learning oriented adaptive beamforming and channel estimation to accomplish the objectives mentioned earlier in ITM 2030.

Further, the MAC layer in 6G will be associated in facilitating the user access to the channel, resource allocation for the spectrum-efficiency, power allocation, cell-free communication, modulation format selection, coding, and decoding scheme selection. Energy- and spectrum-efficiency are essential in THz communication. Therefore, deep learning-based power management algorithms are crucial in enhancing THz communication devices' battery lifetime. However, there is a trade-off between energy-efficiency and spectrum-efficiency; optimizing energy-efficiency leads to reduced spectrum-efficiency and vice-versa. Therefore, ML algorithms are expected to identify the optimal balance between energy-efficiency and spectrum-efficiency.

**ML for routing flow and congestion control**

The deep learning algorithms are instrumental in performing the routing for the wireless network's topology. The intermediate relays' selection is such that the energy-efficiency, spectrum-efficiency, and reliability are higher whereas, the latency is lower. It is also essential to maintain high mobility, ubiquitous connectivity, and survivability, which can be achieved by dynamic deep learning-oriented routing algorithms. Furthermore, intelligent and secure spectrum sharing, and improving the overall network security is essential. Further, the deep learning algorithms reduce the computational complexity, the processing power cost, data size, and storage cost using edge computing.

Finally, we present a consolidated table (see Table 4) which mentions the summary of all the studies in the area of spectrum usage for 6G.

Table 4. Summary of various existing studies with a focus on spectrum utilization.

| Reference | Features |
|---|---|
| [17] | 1. The Federal Communications Commission adopted new rules to encourage the development of new communications technologies in the spectrum above 95 GHz. This spectrum has long been considered the outermost horizon of the usable spectrum range, but rapid advancements in radio technology have made these bands especially ripe for new development.<br>2. There are substantial opportunities for innovation in these frequencies, especially for data-intensive high-bandwidth applications as well as imaging and sensing operations.<br>3. The Commission creates a new category of experimental licenses for use of frequencies between 95 GHz and 3 THz. |
| [21] | 1.Physical layer modelling, Single carrier modulation, constant and near constant envelope modulation is used for Efficient Spectrum utilization (90-200Ghz) hence increase in the wireless network capacity is observed. |
| [33] | 1. Dynamic spectrum management (DSM), is used in order to make full use of the radio spectrum.<br>2. Cognitive radio (CR) is the state-of-the-art enabling technique for DSM.<br>3. With CR, an unlicensed/ secondary user is able to opportunistically or concurrently access spectrum bands owned by the licensed/primary users.<br>4. Blockchain, as an essentially open and distributed ledger, incentivizes the formulation and secures the execution of the policies for DSM<br>5. Artificial intelligence (AI) techniques help the users observe and interact with the dynamic radio environment, thereby improving the efficiency and robustness of CR and blockchain for DSM |
| [34] | Leverage scenarios of THz band Communication are given:<br>1.Local Area Networks: THz networks are suggested to bring seamless transition between THz communication and fiber optics.<br>2.Personal area Networks: THz communication provide data rate like fiber<br>3.Data center Networks: THz links provide promising connectivity at ultrahigh speed in fixed networks.<br>4.Key advancements like Wireless network on chip, Nano network and Inter satellite communication are featured. |
| [36] | 1. Dynamic Spectrum Management via Machine Learning is important for improving spectrum efficiency.<br>2. The state of art research in domain of application of ML in DSD are investigated, categorized and reviewed.<br>3. DSM via machine learning can be divided into four categories: the supervised-learning-based pattern, the unsupervised-learning-based pattern, the semi-supervised-learning-based pattern, and the reinforcement-learning-based pattern.<br>4. There are two operation modes for DSM, namely, the centralized operation mode and the distributed operation mode. |
| [37] | 1.Device-to-device (D2D) communication is a promising technique to improve spectrum efficiency.<br>2. Distributed spectrum allocation framework based on multi-agent deep reinforcement learning is proposed, named Neighbor-Agent Actor Critic (NAAC). |
| [38] | 1. A reliable spectrum sensing scheme for cognitive radio (CR) networks using ML is discussed.<br>2. Spectrum sensing is done using K-nearest neighbor machine learning algorithm.1<br>3. The proposed spectrum sensing scheme aims to improve PU (Primary User) detection capability under varying environments to improve spectral hole detection.<br>4. Mechanisms at both the CR (cognitive radio) level and the FC (fusion center) level ensure reliable spectrum sensing.<br>5. Spectrum sensing scheme outperforms conventional spectrum sensing schemes, both in fading and in nonfading environments, where performance is evaluated using metrics such as the probability of detection, total probability of error, and the ability to exploit data transmission opportunities |
| [43] | 1.Non-orthogonal massively concurrent Access is discussed which seamlessly multiplex large number of users onto available orthogonal resources. New receiving signal model is designed.<br>2.Role of the physical layer in 6G is defined by segregating it in to Low-Phy and High-Phy for data aggregation. |
| [45] | 1.Nitrous oxide is used to illustrate widely tuneable, compact, bright source of THz radiation.<br>2.Gas phase molecular laser based rotational population inversions optically pumped by quantum cascade lasers are suggested. |
| [52] | 1.Creation of high power levels using silicon based integrated technology is discussed. |
| [60] | 1.Receiver SNR reduction issues are sorted out with multiple uniform circular arrays and Dielectric lens antenna methods and Guassian Beams are generated. |
| [74] | 1.Convolutional Neural networks, recurrent neural networks are introduced to serve interactive services with AI techniques.<br>2.Critical path methodology is introduced with RNN to satisfy the requirements like-low latency, flexibility for long shelf life. Programmability and easy usage. |
| [78] | 1.Enhancement of spectrum efficiency with less number of RF channels is done using Huffmann coding based adaptive mode modulation scheme with CSI(Channel state information) .<br>2.Spectrum efficiency is maximized by developing equal probability mode modulation scheme.<br>3.Joint power and probability allocation policy are developed for OAM based verticose Communication.<br>4.AMM(Adaptive mode modulation )EMM(Equal probability mode modulation) features are compared. |
| [91] | 1.The distance varying and frequency selective nature of the THz communication channels is analysed.<br>2.Coherence bandwidth and Significant delay spread are studied.<br>3.Reliable and Ultra high speed wireless communication analysis is proposed in (0.06 to 10THz) band.<br>4.Unified Multi-ray channel modelling based on Ray Tracing Technique which incorporates the propagation model for the line of sight, reflected, scattered and diffracted paths is studied.<br>5.The wideband channel capacity using flat and water filling power allocation strategies are characterized. |
| [92] | 1.Stochastic approach with 300GHz indoor channel modelling methods are introduced to combine both time and frequency domain modelling to show the significant frequency dispersion of ultra-broadband THz channels. |
| [93] | 1.State of art of channel modelling methods and in depth view of SISO and UM-MIMO for THz wireless communication is discussed.<br>2.Deterministic and statistical channel modeling methods are introduced.<br>Sub techniques like Ray Tracing (to analyse the large structures ),finite difference time domain has the ability to resolve small and large scatters on the rough surface of the THz band. |
| [68] | **1**.This paper explores the application of blockchain to radio spectrum management.<br>2. Distributed ledgers offer benefits compared to centralized databases for tracking property rights and assets which could make them effective tools for spectrum management.<br>3.Dynamic and rule-based spectrum use interactions can be tracked and enabled such as: changing the fees for the use of spectrum based on time of day, automatic transfer and reconciliation of spectrum use fees, facilitation of spectrum trading interactions between service providers<br>4. Public Blockchain, Permission Blockchain provides transparent ledger for spectrum management and auditing. |

# IV. 6G Research Activities

The 5G and pre-commercial network specifications were completed in 2019, and currently, 5G is undergoing large scale deployment globally. Simultaneously, to provision the stringent and heterogeneous requirements of the IoE smart services and applications, research activities on 6G have already commenced. According to the market survey, the 6G market will increase at a compound annual growth rate (CAGR) of 70% between the years 2025 and 2030, and is then estimated to touch 4.1 Billion US dollars by the year 2030 [81]. Also, in comparison to other 6G components required for edge and cloud computing and AI, the infrastructure necessary for communication will offer the largest market share of approximately 1 billion US dollars. Further, the critical components of 6G, i.e., AI chipsets, are expected to amount to about 250 million units by 2028 [81].

Even though the 6G research is at an infant stage, studies are being conducted worldwide in regard to the standardization of 6G [82-90]. Various research communities/organizations/governments have initiated the projects related to 6G of which, the prominent ones are as follows:

1. The joint university microelectronic project (JUMP), which was launched by the Defence Advanced Research Projects Agency (DARPA) in 2016, resulted in the FCC to grant the frequency band between 95GHz to 3THz for research on 6G. This also led to the U.S.A set the pace in regard to the research on 6G [7].

2. In 2017, the European Union launched the Terapod and Terranova projects, which were sponsored by 5GPPP Phase 1, which were then followed by more projects in Phase 2 and Phase 3 [7]. Further, in 2018, the Chinese Communication Standardization Association (CCSA) launched two projects focusing on the vision and requirements of 6G and defining the key technologies for 6G, respectively [7].

3. The 6G Flagship research program, sponsored by the Academy of Finland and commenced by the University of Oulu, was initiated in 2019 to conduct the co-creation of an ecosystem for 6G innovation 5G adoption. The vision of the 6G flagship program is a data-driven society with unlimited, instant wireless connectivity. Research under 6G Flagship is sub-divided into four unified planned categories viz., wireless connectivity, distributed computing, services, features, and applications [82]. Initially, VTT Technical Research Center of Finland Ltd., Oulu University of Applied Sciences, Nokia, Business Oulu, and Aalto University were the collaborators. InterDigital and Keysight Technologies joined the program at a later stage. Further, in 2019, an agreement was signed between the South Korean government and the University of Oulu, Finland, to develop 6G technology [82].

4. Samsung Electronics has initiated an R & D center to develop primary technologies for 6G mobile networks. To stimulate the growth of solutions and the 6G's standardization, Samsung is conducting extensive research on cellular and networking technologies. The company has enhanced the next-generation telecommunication research team to a center and has also published white papers in 2020, providing insight into the key technologies and enablers for 6G [16].

5. Further, LG has established the first research laboratory at the Korea Advanced Institute of Science and Technology to conduct 6G research activities [84]. SK Telecom has also initiated joint research on 6G in collaboration with Samsung, Nokia, and Ericsson [85].

6. In Japan, a 2 billion US dollar support industry has been initiated for research on 6G technology with NTT and Intel deciding to form a partnership for researching 6G mobile network technology [86].

7. 6G research has also started in China, as officially announced by the Ministry of Science and Technology [87]. Furthermore, Huawei, the Chinese vendor, has already started 6G research at its research center in Ottawa, Canada [88].

8. Also, the NYU WIRELESS research center, which includes approximately 100 faculty members and graduate students, is working on communication foundations, machine learning techniques, quantum nano-devices, and primary 6G testbeds [89].

9. From academia, the 6G research began in 2019 with the first 6G Wireless Summit, which was held in Levi, Finland [80]. In addition, other mini-workshops and conferences have also been conducted globally to examine the prospect of 6G, including the Huawei 6GWorkshop, the Wi-UAV Workshop of Globecom 2018, and the Carleton 6GWorkshop [90].

10. IEEE began the IEEE Future Network with the tagline viz., 'Enabling 5G and beyond' in 2018. Further, the ITU-T Study Group 13 also established the ITU-T Focus Group Technologies for Network 2030, intending to understand future network service requirements by 2030 [7].

# V. Issues and Challenges

In this section, we present the various challenges and issues which exist in the research over 6G, which are as follows:

1. The spectrum resources in 6G will be scarce due to which it will be required to manage the spectrum efficiently. This will necessitate the investigation and implementation of efficient spectrum-sharing mechanisms and novel spectrum management techniques to ensure the desired service quality simultaneously ensuring maximum resource utilization. Hence, the subsequent issues necessitate being addressed: (i) how spectrum can be shared and managed considering the heterogeneity of the 6G communications, and (ii) the cancellation of interference via existing or novel interference cancellation techniques. With the deployment of 6G, high data rates in the range of Tbps will be desired; however, the THz band is severely affected by the propagation loss and atmospheric absorption, limiting the long-distance THz communication. Therefore, novel transceiver design/architecture will be required, which will enable the transceivers to operate at THz frequencies simultaneously, ensuring the efficient utilization of the complete available bandwidth. Further associated challenges concerning the aforementioned exist in the design of antennas operating at THz frequency, providing minimum gain with the effective area, and the concerns related to health and safety.

2. A major challenge is in regard to channel modelling of the THz band due to its being subjected to the atmospheric conditions. Therefore, due to climatic conditions' highly unpredictable nature, channel modelling of the THz band assumes complexity and requires much investigation [91]. Specifically, the THz spectrum band incurs a frequency-based path loss owing to the molecular absorption, which motivates the development of distance adaptive and multiple wideband design solutions for the physical layer. Also, the coherence bandwidth in the THz band is known to (i) vary inversely proportional to the delay spread, (ii) increase with the operating frequency, and (iii) decrease with the transmission range [92]. Therefore, it is required to develop an enabling technology in this regard to increase the coherence bandwidth. Further, the THz band's frequency selective nature results in the broadening of the signals in the time domain, which in turn limits the minimum separation between consecutive transmissions in view of avoiding inter-symbol interference. Therefore, the variation of such broadening with respect to operating frequency and transmission distance needs more investigation. Also, for the ultra-massive multiple inputs multiple output THz systems, the spatial degrees of freedom amount are defined as the maximum spatial multiplexing gain that an ultra-massive multiple inputs multiple output THz system could support [93]. Since the spatial degree of freedom varies with the array area, array geometry, and angular spread, how the spatial degrees of freedom can be increased needs to be investigated. In regard to the capacity, for a single input single output THz channel, the capacity depends on the path gain and can be estimated by dividing the THz spectrum into multiple sub-bands; however, for an ultra-massive numerous inputs multiple output THz channel, the capacity depends on the path again, the spatial degrees of freedom, beamforming, and spatial multiplexing techniques employed on the THz antenna array. The aforementioned issues need more investigation. Lastly, 6G networking will be complicated due to the 3D in nature, which implies adding a new dimension and an extension vertically. Therefore, novel methods to manage the resources need to be developed in addition to advanced techniques for supporting mobility, routing protocol, multiple access, and scheduling.

3. With 6G operating at THz frequency, the wireless coverage ranges of a 6G access point (AP) will be much minimized but, it will be impossible to split the cell indefinitely as this starts to approach the limit of improvement in the system capacity. Further, such a network architecture will start to demonstrate ineffectiveness when the size of the cell and the distance of transmission become smaller owing to the amount of APs equalling or exceeding the users' terminals amount. Hence, this requires the dawn of the concept of a network provisioning the user. For such cases, the de-cellular network architecture can be adopted, which enables the organization of dynamic APs Group (APG) in view of seamlessly provisioning every user without involving the user, simultaneously letting the user believe that the network follows the user's communication surroundings through intelligence and also organizes the APG and resources needed by the user flexibly [94]. As an extension, the User-Centric Ultra-Dense Network (UUDN) has also been proposed, which is a wireless network with similar densities of AP and the users' density [7]. Although de-cellular and UUDN fit the higher frequency range operation, there exist many challenges before the practical implementation viz., (a) The realization of serving the user in a dynamic network is challenging since this involves the APs to cooperate and transmit with each other, which requires the design of novel network architecture that caters to the exact needs of the user, supports uninterrupted communication during the users' movement, and enhances spectrum- and energy-efficiency simultaneously with the user experience, (b) To manage the mobility of users, novel methods need to be explored which will ensure the dynamic adaptation of the AP group of any user constantly on the move, (c) It will be mandatory to incorporate AI techniques within the de-cellular network and UUDN to manage the complex interference issues efficiently. Advanced methods in this regard need more investigation, (d) Further, as opposed to the current manner of resource allocation based on a single cell, advanced methods need to be developed which are based on the users and the corresponding AP group, and (e) With the dynamic nature of de-cellular network and UUUDN, security mechanisms between the APs and the AP groups assumes importance. This will mandate trusted authentication and secure transmission to be established between the APs, the user to the AP, and the AP groups. Also, since the AP group, which serves a user, will be dynamically generated, it will be required to identify the AP group's security protocol when an AP enters or exits.

4. The connectivity at the backhaul in 6G will require very high capacity due to the high density of the access networks, supporting heterogeneous data from diverse users located at various locations. Therefore, the backhaul network of the 6G will be required to handle enormous amounts of data between the access and the core network. Currently, the optical fiber and free-space optics are the proposed solutions concerning providing high capacity backhaul networks in 6G; however, any further enhancement with the exponentially increasing users' data and demands in 6G will mandate the investigation of advanced networks.

5. The 6G networks will focus on human-centric communication due to which the aspects of privacy, secrecy, and security will have to be ensured through advanced AI technologies and Big Data. Therefore, novel physical layer methods need to be developed for 6G communication, which will ensure that interactions with the upper layers can be conducted securely. Currently, quantum key distribution and advanced ML techniques are proposed to provide automated security; however, much research is desired in this area.

6. A trade-off aspect in 6G communications will be that of security and spectral-efficiency since higher end-to-end communication without any attacks consumes higher bandwidth, reducing the available bandwidth for data transmission. Therefore, there is a requirement to develop schemes that provide acceptable security and spectral-efficiency, which requires further investigation on AI and encryption algorithms. Another trade-off will be that of spectral-efficiency and energy-efficiency, which can be relaxed in 6G communications using the energy harvesting technology. The consumer nodes can harvest the radio, vibratory, and solar energy from the ambient environment by addressing energy utilization complications. In turn, this will be essential in reducing the spectrum- and energy-efficiency trade-off concerning the ever-changing radio propagation conditions.

7. The 5G technology enabled the network to provide and share the services using network slicing, which uses software-defined networking and network function virtualization [2]. However, network slicing may not be efficient for highly heterogeneous and complex network scenarios such as 6G. Therefore, there is a need to convert the 5G enabled network service to the 6G network intelligent service, which will adjust the different parameters adaptively, endeavouring enhanced performance.

8. The 6G communication will be required to provide remarkably low latency and energy consumption and ensure highly scalable and reliable systems. The existing blockchain methods will pose extreme limitations in reliability, scalability, latency, and energy consumption. Therefore, regarding the aforementioned, it will be necessitated to design advanced blockchain technologies

Finally, in Table 5, we suggest the various directions which can be adopted in response to the existing research challenges on 6G.

Table 5. Various existing challenges in research on 6G and possible direction to obtain the solutions.

| Sl. No. | Challenges | Directions |
|---|---|---|
| 1 | • Spectrum (Resource) Sharing and Managing considering heterogeneous traffic.<br>• Interference cancellation | • M- MIMO Fixed phase shifter group connected<br>• Index Modulation (IM) to increase the system spectral efficiency<br>• Carrier aggregation |
| 2 | • Novel transceiver design/architecture<br>• Design an antenna operating at THz | • The 6G transceiver parameters can be adaptively tuned by Machine learning algorithms.<br>• Electronics- and photonics-based approach, considering the transmitting power and power-efficiency, complexity, cost, and size<br>• Semiconductor technology and meta-materials, e.g., graphene-based electronics<br>• Deep Q-learning and Federated Learning based transceiver design<br>• Reconfigurable intelligent surface (RIS)<br>• Analog beamforming, hybrid precoding, and delay-phase precoding |
| 3 | • Channel modeling of THz Band | • Constant or near-constant envelope modulation (CPM and (D)QPSK)<br>• SC-IM approaches are a balanced trade-off between system performance/power efficiency and hardware cost/detector computational complexity<br>• Single carrier for lower cubic metric (CM)<br>• Machine learning and AI based, model free data driven learning and optimization techniques. |
| 4 | • Dynamic network service provisions | • Dynamical convergence can be efficient through designing a software-defined and re-configured radio access network architecture, |
| | • Development of Advance methods based on users and corresponding AP group | so that the communication, data caching, and computing can be dynamically re-organized.<br>• Merging xRAN forum and C-RAN alliance to formally define the requirements that would help achieve these challenges<br>• Design de-cellular architecture or User-Centric Ultra-Dense Networks and make use of dynamic AP grouping method. |
| 5 | • Advanced techniques to support maximum users | • e-MBB enhanced Mobile Broadband service |
| 6 | • Development of physical layer methods to ensure secured interaction | • Cybersecurity and Block chain-based secured service brokering schemes |
| 7 | • Conversion of 5G enabled network service to the 6G network | • **Physical layers**:AI-based technologies, including deep neural networks, K-Means, and supervised/unsupervised learning, which could be used in different physical layers to predict traffic and enhance security.<br>• **Network Architecture**: AI based edge security via security systems and fine-grained controls. Specific deep learning techniques to detect threats in edge computing<br>• LDPC and CRM coding schemes capable of enhancing the security of the transmitted data.<br>• DNA-cryptography techniques are new authentication mechanisms to protect data security and privacy. |
| 9 | • Necessity to design advanced Block chain techniques to address scalability, reliability, latency, and energy consumption | • **Latency**: contextually agile e-MBB,<br>• **Reliability**: Event defined URLLC-ultra reliable low-level communication<br>• **Performance**: COC: Computer oriented communication services. |

# VI. Conclusion

In this article, we surveyed the most recent advances in enabling the 6G technology. We established that, in the realization of 6G, there is a corresponding technological trend for tackling every challenge. Among the many directions, in this survey article, we focused on the ones which tackle the capacity issues and are the possible solutions to enhance the spectrum-efficiency in 6G communications. We presented these trends in detail and then identified the challenges which need to be addressed before the practical deployment for realizing 6G communications. From our detailed survey, we concluded that the 6G enabled systems would unlock the unlimited potential of the next-generation advanced applications via the enabling of the Internet of Everything. Also, the advanced AI techniques will form an integral part of the 6G system in view of solving the complex network optimization problems. Further, the THz frequency range will be used for communication in 6G, which will require the proposal of novel models for the THz communication. We also presented the international research activities that aim to create a vision for future 6G communications. Lastly, we concluded with several challenges and essential guidelines and identified that 6G would support a Ubiquitous Intelligent Mobile Society with intelligent life and industries.